\documentstyle[aas2pp4,psfig]{article}

\newcommand\mdot   {\hbox {${\dot M}$}}
\newcommand\sz     {S$_{\rm z}$}

\newcommand\pp     {$\pm$}

\newcommand\micros  {$\mu$s}

\righthead{KiloHertz Quasi-Periodic Oscillations in Cygnus X-2}
\slugcomment{Submitted to ApJ Letters, October 1997}

\begin{document}

\title{Discovery of KiloHertz Quasi-Periodic Oscillations in the Z
source Cygnus X-2}

\author{Rudy Wijnands\altaffilmark{1},
        Jeroen Homan\altaffilmark{1},
        Michiel van der Klis\altaffilmark{1},
	Erik Kuulkers\altaffilmark{2},
	Jan van Paradijs\altaffilmark{1,3},
	Walter H. G. Lewin\altaffilmark{4},
	Frederick K. Lamb\altaffilmark{5},
	Dimitrios Psaltis\altaffilmark{6},
        Brian Vaughan\altaffilmark{7}
        }

\altaffiltext{1}{Astronomical Institute ``Anton Pannekoek'',
University of Amsterdam, and Center for High Energy Astrophysics,
Kruislaan 403, NL-1098 SJ Amsterdam, The Netherlands;
rudy@astro.uva.nl, homan@astro.uva.nl, michiel@astro.uva.nl,
jvp@astro.uva.nl}

\altaffiltext{2}{Astrophysics, University of Oxford,
Nuclear and Astrophysics Laboratory, Keble Road, Oxford OX1 3RH,
United Kingdom; e.kuulkers1@physics.oxford.ac.uk}

\altaffiltext{3}{Departments of Physics,
University of Alabama at Huntsville, Huntsville, AL 35899}

\altaffiltext{4}{Department of Physics and Center for Space Research,
Massachusetts Institute of Technology, Cambridge, MA 02139;
lewin@space.mit.edu}

\altaffiltext{5}{Departments of Physics and Astronomy,
University of Illinois at Urbana-Champaign, Urbana, IL 61801;
f-lamb@uiuc.edu}

\altaffiltext{6}{Harvard-Smithsonian Center for Astrophysics, 60
Garden St., Cambridge, MA 02138; demetris@cfata1.harvard.edu}

\altaffiltext{7}{Space Radiation Laboratory, California
Institute of Technology, 220-47 Downs, Pasadena, CA 91125;
brian@thor.srl.caltech.edu}

\begin{abstract}

During observations with the Rossi X-ray Timing Explorer from June
31st to July 3rd 1997 we discovered two simultaneous kHz QPOs near 500
Hz and 860 Hz in the low-mass X-ray binary and Z source Cygnus X-2. In
the X-ray color-color diagram and hardness-intensity diagram (HID) a
clear Z track was traced out, which shifted in the HID within one day
to higher count rates at the end of the observation. Z track shifts
are well known to occur in Cygnus X-2; our observation for the first
time catches the source in the act.  A single kHz QPO peak was
detected at the left end of the horizontal branch (HB) of the Z track,
with a frequency of 731\pp20 Hz and an amplitude of
4.7$^{+0.8}_{-0.6}$\% rms in the energy band 5.0--60 keV.  Further to
the right on the HB, at somewhat higher count rates, an additional
peak at 532\pp43 Hz was detected with an rms amplitude of
3.0$^{+1.0}_{-0.7}$\%. When the source moved down the HB, thus when
the inferred mass accretion rate increased, the frequency of the
higher-frequency QPO increased to 839\pp13 Hz, and its amplitude
decreased to 3.5$^{+0.4}_{-0.3}$\% rms. The higher-frequency QPO was
also detected on the upper normal branch (NB) with an rms amplitude of
1.8$^{+0.6}_{-0.4}$\% and a frequency of 1007\pp15 Hz; its peak width
did not show a clear correlation with inferred mass accretion rate.
The lower-frequency QPO was most of the time undetectable, with
typical upper limits of 2\% rms, no conclusion on how this QPO behaved
with mass accretion rate can be drawn.  If the peak separation between
the QPOs is the neutron star spin frequency (as required in some
beat-frequency models) then the neutron star spin period is 2.9\pp0.2
ms (346\pp29 Hz). This discovery makes Cygnus X-2 the fourth Z source
which displays kHz QPOs. The properties of the kHz QPOs in Cygnus X-2
are similar to those of other Z sources.

Simultaneous with the kHz QPOs, the well known horizontal branch QPOs
(HBOs) were visible in the power spectra. At the left end of the HB
the second harmonic of the HBO was also detected.  We also detected
six small X-ray bursts.  No periodic oscillations or QPOs were
detected in any of them, with typical upper limits of 6--8 \% rms.

\end{abstract}


\section{Introduction \label{intro}}

The low-mass X-ray binary (LMXB) and Z source (Hasinger \& van der
Klis 1989) Cygnus X-2 is one of the best studied LMXBs. In the X-ray
color-color diagram (CD) a Z shaped track is traced out. Its branches
are called horizontal branch (HB), normal branch (NB), and flaring
branch (FB). The rapid X-ray variability in Z sources is closely
related to the position of the sources on the Z. Quasi-periodic
oscillations (QPOs) are present on the HB and the upper NB with a
frequency between 18 and 60 Hz (called HBO). QPOs with a frequency
between 5 and 20 Hz are present on the lower part of the NB and on the
beginning of the FB (called N/FBOs).  Both the X-ray spectral changes
and the rapid X-ray variability are thought to be due to changes in
the mass accretion rate (\mdot) (e.g. Hasinger \& van der Klis 1989;
Lamb 1991), which is inferred to be lowest on the HB, increasing along
the Z and reaching the Eddington mass accretion rate at the NB/FB
vertex.

Recent studies of Cygnus X-2 (EXOSAT: Kuulkers et al. 1996; Ginga:
Wijnands et al. 1997a) show that Cygnus X-2 displays motion of the Z
track in the hardness-intensity diagram (HID). Following Kuulkers et
al. (1996), we call the different positions of the Z track in the HID
different ``overall intensity levels''.  When the overall intensity
level changes, the morphology of the Z changes and the whole Z moves
in the CD as well as in the HID.  Wijnands et al. (1997a) found that
the X-ray timing behavior on the lower part of the NB differs
significantly for different overall intensity levels.  The fact that
the X-ray spectral and X-ray timing behavior change with changing
overall intensity level suggests that something else in addition to
\mdot\, determines the Z track and the timing properties of Cygnus
X-2.  A precessing accretion disk, strongly suggested by the 78 day
period found in Cygnus X-2 using the ASM/RXTE long-term X-ray light
curve (Wijnands, Kuulkers, \& Smale 1996), is a possible explanation
of the long-term changes.

In numerous LMXBs, mostly atoll sources (Hasinger \& van der Klis
1989), kHz QPOs have been observed (see van der Klis 1997 for a recent
review on kHz QPO properties).  So far, kHz QPOs have been found in
three Z sources (Sco X-1: van der Klis et al. 1996a, 1997; GX\,5$-$1:
van der Klis et al. 1996b; GX\,17$+$2: Wijnands et al. 1997b). In this
letter, we report the discovery of two simultaneous kHz QPOs in the Z
source Cygnus X-2.

\section{Observations and Analysis  \label{observations}}

We observed Cygnus X-2 with the Rossi X-ray Timing Explorer (RXTE)
from June 30th until July 3th 1997. A total of 108 ksec of data was
obtained. Data were collected simultaneously with 16 second time
resolution in 129 photon energy channels (effective energy range 2--60
keV), and with 122 \micros\, time resolution in two channels (2--5.0
keV, 5.0--60 keV).  A CD and HID were made using the 16 s data, and
power density spectra were calculated from the 122 \micros\, data
using 16 s data segments.

For measuring the kHz QPO properties we fitted the 128--4096 Hz power
spectra with a constant plus a broad sinusoid representing the
deadtime-modified Poisson level (Zhang 1995, Zhang et al. 1995), and
one or two Lorentzian peaks representing the kHz QPOs.  For measuring
the HBO we fitted the 8--256 Hz power spectra with a constant plus a
power law representing the continuum, and one or two Lorentzian peaks
representing the HBO and its second harmonic. Differential dead time
effects (van der Klis 1989) were negligible due to the relatively low
count rates. The errors on the fit parameters were determined using
$\Delta\chi^2 = 1.0$; upper limits were determined using $\Delta\chi^2
= 2.71$, corresponding to 95\% confidence levels.  The kHz QPOs were
only detected in the 5.0--60 keV band. In the 2--5.0 keV and the
combined 2--60 keV energy bands the kHz QPOs were undetectable.
Therefore, we used the 5.0--60 keV band throughout our analysis.

For the CD we used for the soft color the logarithm of the
count rate ratio between 3.5--6.4 keV and 2.0--3.5 keV, and for the
hard color the logarithm of the ratio between 9.7--16.0 keV and
6.4--9.7 keV.  For the HID we used as intensity the logarithm of the
count rates in the energy band 2.0--16.0 keV, and for the hardness we
used the hard color from the CD. For $\sim$12\% of the time detector 4
or 5 was off. For the CD and HID we only used the data of the three
detectors which were always on in order to get the position of Cygnus
X-2 on the Z during the whole observation.  For the power density
spectra we used all available data.

We used the \sz\, parameterization (see Wijnands et al. 1997a and
references therein) for measuring the position along the Z (HB/NB
vertex at \sz=1, NB/FB vertex at \sz=2). Due to the fact that the
position on the Z could be much better determined in the HID than in
in the CD we used the \sz\, parameterization for the Z track in the
HID. Thus the \sz\, values quoted in this paper represent the position
of the source on the Z track in the HID. By using logarithmic values
for the colors and the intensity, \sz\, does not depend on the values
of those quantities but only on their variations (Wijnands et
al. 1997a).

\section{Results \label{results}}

\begin{figure}[t]
\begin{center}
\begin{tabular}{c}
\psfig{figure=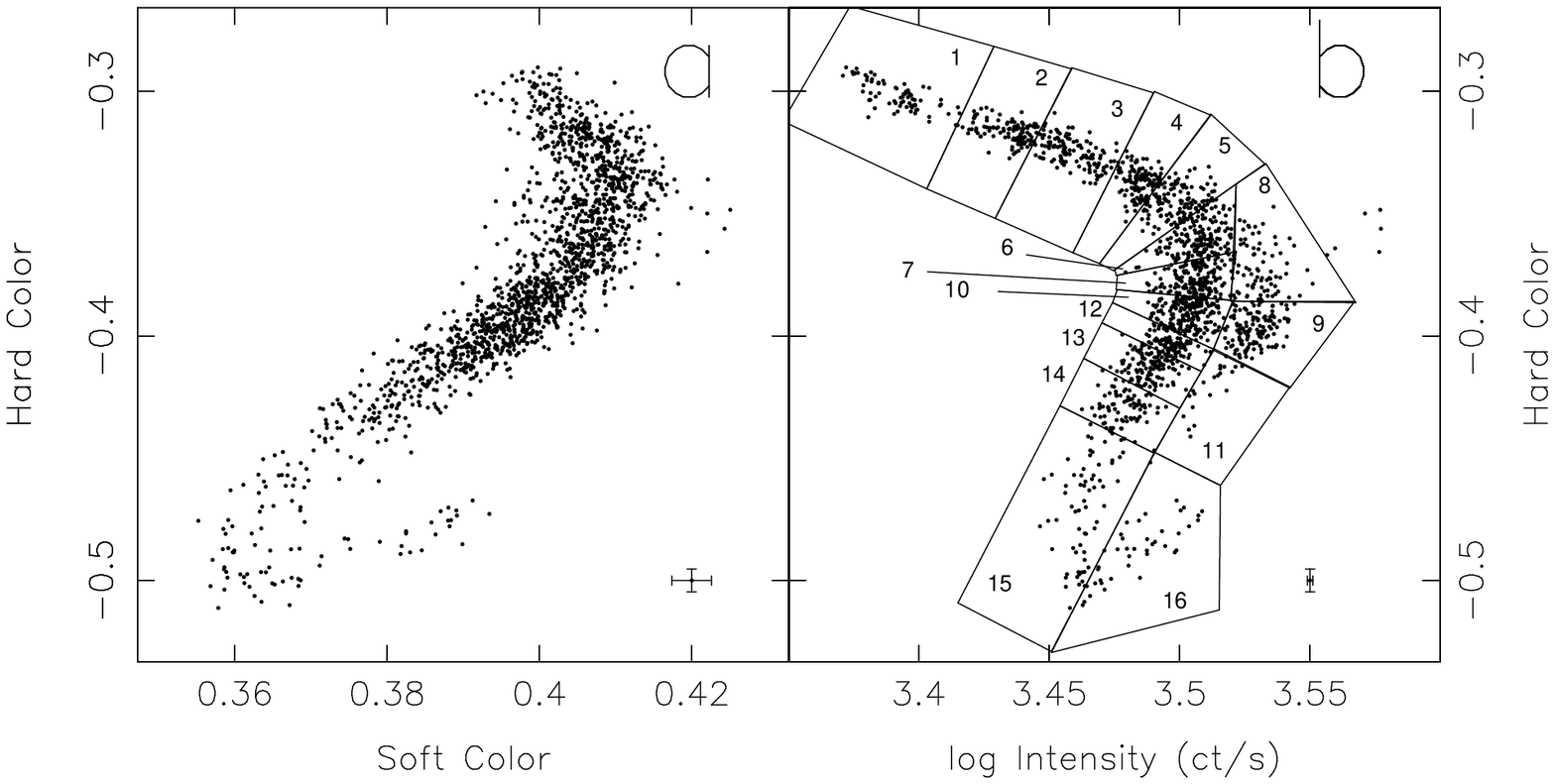,width=18cm}
\end{tabular}
\caption{Color-color diagram ({\it a}) and
hardness-intensity diagram ({\it b}) of Cygnus X-2. The soft color is
the logarithm of the count rate ratio between 3.5-6.4 keV and 2.0-3.5
keV; the hard color is the logarithm of the count rate ratio between
9.7-16.0 keV and 6.4-9.7 keV; the intensity is the logarithm of the
3-detector count rate in the photon energy range 2.0--16.0 keV. Both
diagrams are corrected for background ($\sim$50 cts/s in the energy
range 2.0--16.0 keV); the count rate is not dead-time corrected
(4--6\% correction). All points are 64 s averages. Typical
error bars are shown at the bottom right corners of the diagrams. The
boxes in the HID are the regions which were selected to study the timing
behavior.  \label{CD_HID}}
\end{center}
\end{figure}

In Fig.\,\ref{CD_HID} the CD and the HID are shown. Cygnus X-2 traced
out an almost complete Z track. The HID (Fig. \ref{CD_HID}b) clearly
shows that part of the data near the HB/NB vertex (regions 8, 9 and 11
in Fig. \ref{CD_HID}b) was shifted to higher count rate with respect
to the rest. No sudden jump in count rate is visible in the X-ray
light curve so Cygnus X-2 moved smoothly to higher count rates. At the
end of the observation the source moved in 0.7 days from the HB/NB
vertex to the left of the HB and back to the vertex again which by
then had shifted to higher count rates.  This is the first time we
catch the source in the act.  Previously, shifted Z's were only
detected on time scales of weeks to months (between observations), not
during one observation.  In the CD (Fig. \ref{CD_HID}a) the shift is
not visible due to the fact that this shift takes place parallel to
the NB (see also Wijnands et al. 1997a) and therefore all data
overlap. Also, the error bars in the CD are large, giving larger
uncertainties than in the HID.  Some of the timing properties have
previously been observed to change when the Z track shifts in the CD
and HID (Wijnands et al. 1997a). For the small shift in the Z track in
our observation these changes were undetectable for the QPOs discussed
in this paper.

\begin{figure}[t]
\begin{center}
\begin{tabular}{c}
\psfig{figure=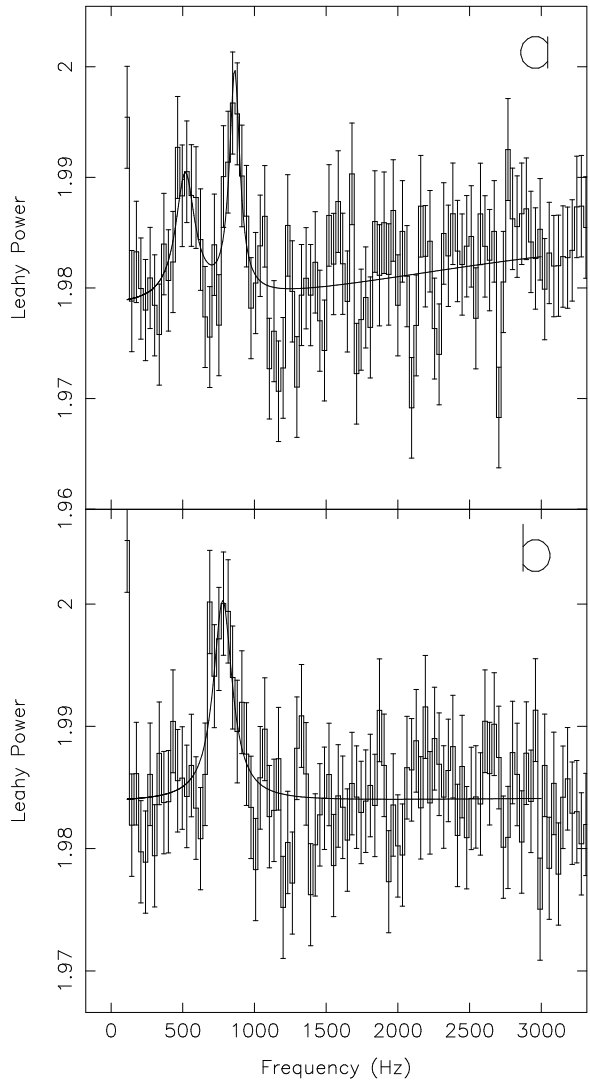,width=8cm}
\end{tabular}
\caption{Typical Leahy normalized power spectra
in the energy range 5.0--60 keV showing the kHz QPOs.  The upwards
slope at kHz frequencies is due to instrumental dead time. (a) is the
power spectrum for the interval 2 July 1997 03:36--08:19 UT; (b) the
power spectrum for the interval 2 July 1997 09:37--12:57 UT.
\label{powerspectra}}
\end{center}
\end{figure}

Between 2 July 1997 03:36 and 08:19 UT (5.8 ksec exposure time) we
detected two simultaneous kHz QPOs (Fig.\,\ref{powerspectra}a) with
frequencies of 516\pp27 Hz and 862\pp11 Hz (peak separation of
346\pp29 Hz), at 3.4$\sigma$ and 4.4$\sigma$ significance
respectively. The FWHM and rms amplitude were 170$^{+66}_{-51}$ Hz and
3.6$^{+0.6}_{-0.5}$\% for the lower-frequency QPO, and
94$^{+31}_{-25}$ Hz and 3.5\pp0.4\% for the higher-frequency
QPO. Upper limits to the rms amplitude of the QPOs for the same time
interval in the energy range 2--5.0 keV were 2.4\% and 1.1\% rms,
respectively. Simultaneously with these kHz QPO we detected the
HBO. Between 2 July 1997 09:37 and 12:57 UT (7 ksec exposure time) we
detected a single kHz QPO (Fig. \ref{powerspectra}b) with a frequency
of 779\pp16 Hz at 4.9$\sigma$ significance, with a FWHM of
177$^{+52}_{-40}$ Hz, and an amplitude of 4.7$^{+0.6}_{-0.5}$\%
rms. The upper limit on the amplitude of the QPO in the 2--5.0 keV
band was 2.7\% rms. kHz QPOs were detected at several other times,
although usually below a 3$\sigma$ significance level.

\begin{figure}[t]
\begin{center}
\begin{tabular}{c}
\psfig{figure=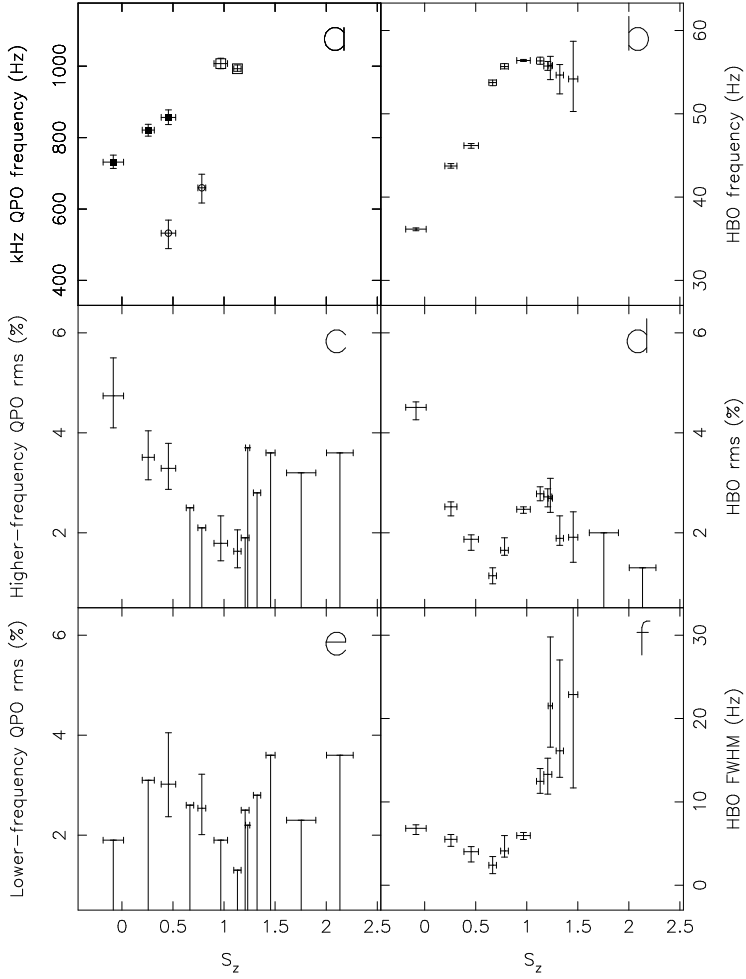,width=10cm}
\end{tabular}
\caption{(a) Frequency of the kHz QPOs, (b) frequency of
the HBO fundamental, (c) the rms amplitude of the higher-frequency kHz
QPO, (d) the rms amplitude of the HBO fundamental, (e) the rms
amplitude of the lower-frequency kHz QPO, and (f) the FWHM of the HBO
fundamental as a function of \sz. In (a) the squares represent the
higher-frequency QPO and the circles the lower-frequency QPOs. Open
symbols represent detections of the kHz QPOs with significance between
a 2 and 3$\sigma$, the filled points represent detections with
$>$3$\sigma$ significance.
\label{qpoversussz}}
\end{center}
\end{figure}

The Z track in the HID was not homogeneously covered. In order to get
at all positions on the Z track enough statistics to detect the kHz
QPOs or determine useful upper limits we selected power spectra in the
regions of the HID indicated in Fig. \ref{CD_HID}b. Afterwards we
determined the average \sz\, value of the power spectra selected.  The
frequency and the rms amplitudes of the kHz QPOs versus \sz\, are
shown in Fig. \ref{qpoversussz}.  At the left most end of the HB
(lowest count rates) only the higher-frequency QPO could be
detected. When the source moved to somewhat higher count rates a
second, lower-frequency kHz QPO became detectable. Further to the
right on the HB the lower-frequency kHz QPO was undetectable again
with upper limits of 2--3\%. Near the HB/NB vertex the
higher-frequency QPO was detected, but further down the NB the
higher-frequency peak became undetectable, with upper limits of 2--4\%
rms.  With increasing \mdot\, the frequencies of the higher-frequency
kHz QPOs increased and its amplitude decreased
(Fig. \ref{qpoversussz}a and c). The FWHM of the peak did not show a
clear correlation with \mdot. Due to the small number of detections of
the lower-frequency QPO no conclusions can be drawn about the behavior
of this QPO with \mdot, although there are indications that its
frequency also increases with \mdot\, (see Fig. \ref{qpoversussz}a).
It was not possible to determine if the kHz QPO properties changed
significantly as the Z track moved in the HID (regions 6, 7, and 12
versus 8 and 9 in Fig. \ref{CD_HID}b). This was also true for the
properties of the HBO.

Simultaneously with the kHz QPOs we detected the HBO. Its properties
(in the 5.0--60 keV band) versus \sz\, are also shown in
Fig. \ref{qpoversussz}. The HBO second harmonic was only detectable in
region 1 (\sz=$-$0.09\pp0.1) of Fig.\,\ref{CD_HID}b with a frequency of
73.6\pp0.8 Hz, an rms amplitude of 3.1$^{+0.2}_{-0.4}$\%, and a FWHM
of 13$^{+2}_{-3}$ Hz. The HBO fundamental in this region had a frequency
of 36.2\pp0.2 Hz, an rms amplitude of 4.5$^{+0.1}_{-0.3}$\%, and a
FWHM of 6.8$^{+0.4}_{-0.7}$ Hz. When the source moved further to the
right on the HB the second harmonic became undetectable with typical
upper limits on the amplitude of 1--2\% rms.  The HBO fundamental was
detected down to about halfway the NB (region 14; \sz=1.46\pp0.05). Its
frequency increases from 36.2\pp0.2 Hz at the left end of the HB to
56.4\pp0.1 Hz at the HB/NB vertex (Fig. \ref{qpoversussz}b). After
that its frequency remained approximately constant. The rms amplitude
of the HBO first decreased on the HB with increasing \mdot, but around
\sz=0.6 it started to increase again (Fig. \ref{qpoversussz}d). At
around \sz=1.1 it decreased again with increasing \mdot\, until it
became undetectable further down the NB. Its FWHM also first decreased
on the HB with increasing \mdot\, (Fig. \ref{qpoversussz}f), and at
the same position on the Z where the amplitude of the HBO started to
increase again with increasing \mdot, the FWHM also started to
increase again.

We detected six bursts in the X-ray light curve (typical durations of
4-6 s), which are very similar to those found in EXOSAT (Kuulkers, van
der Klis, \& van Paradijs 1995) and Ginga (Wijnands et al. 1997a)
data. We searched for periodic oscillations and QPOs during the
bursts, but none were detected (typical upper limits of 6--8\% rms). A
more detailed study of the X-ray bursts will be reported elsewhere.

\section{Discussion \label{discussion}}

We have discovered two simultaneous kHz QPOs near 500 Hz and 860 Hz in
the Z source Cygnus X-2. The frequency of the higher-frequency QPO
increased when the source moved from the left end of the HB to the
upper NB, and there are indications that the same is true for the
frequency of the lower-frequency QPO.  Cygnus X-2 is the fourth Z
source which displays kHz QPOs. Most likely the remaining two Z
sources (GX 340$+$0 and GX 349$+$2) will also display kHz QPOs when
they are observed on the (left end of the) HB (note that GX 349$+$2
has never been observed in the HB).

The properties of the kHz QPOs in Z sources are very similar to each
other. In each source the kHz QPOs increase in frequency and the
higher-frequency QPO decreases in amplitude when the sources move down
the Z, thus when \mdot\, increases.  The amplitude of the
lower-frequency QPO and the FWHM of the peaks do not have a clear
correlation with \mdot. The kHz QPOs in Z sources are also very
similar to what is observed (e.g. Strohmayer et al. 1996; Zhang et
al. 1996; Berger et al. 1996; Smale et al. 1997) in other, less
luminous LMXBs (the atoll sources [Hasinger \& van der Klis 1989]).
In both type of sources the frequency of the kHz QPOs increases with
inferred \mdot, the kHz QPOs are strongest in the highest photon
energy band, the maximum frequencies so far detected are between 1000
and 1200 Hz, and the separation between the kHz QPOs lies between 250
and 370 Hz (see van der Klis 1997 for a recent review on kHz QPO
properties).  Therefore, it seems likely that kHz QPOs in atoll
sources and in Z sources are produced by the same mechanism.

In Sco X-1 (van der Klis et al. 1996a; 1997) the peak separation
between the two kHz QPOs decreases with increasing
\mdot\footnote{Recently, M\'endez et al. (1997) found that also in
the atoll source 4U 1608$-$52 the peak separation may not be
constant.}. In GX 17$+$2 a similar decrease in peak separation could
not be excluded (Wijnands et al. 1997b) and due to the sparcety of
detections with two simultaneous kHz QPO in the present data nothing
can be said for Cygnus\,X-2. If the peak separation was a measure for
the neutron spin frequency then the spin frequency in Cygnus X-2 is
346\pp29 Hz (spin period of 2.9\pp0.2 ms). In the four Z sources for
which kHz QPOs have been discovered the kHz QPOs were detected
simultaneously with the HBO, ruling out the possibility that both the
HBOs and the kHz QPOs are magnetospheric beat frequencies (van der
Klis et al. 1997b).

At the left end of the HB the amplitude, in the energy range 5--60
keV, of the higher-frequency QPO in Cyg X-2 is 4.7$^{+0.8}_{-0.6}$\%
rms.  This amplitude is smaller than the typical amplitudes of the kHz
QPOs in the atoll sources, which are thought to have significantly
weaker magnetic fields, as predicted by the sonic-point model (Miller,
Lamb, \& Psaltis 1997). The QPOs in Cyg X-2 are similar in strength to
those in GX 17+2 (2--5\% rms in 5--60 keV), even though GX 17+2 may
have a weaker magnetic field, as suggested by its weaker HBO and by
X-ray spectral modeling (Psaltis, Lamb, \& Miller 1995).  This
indicates that if the strength of the stellar magnetic field affects
the amplitudes of the kHz QPOs, it is not the only variable that does
so, and that other variables, such as the multipolar structure and
orientation of the magnetic field, and the spin rate of the star, also
play a role. If the orientation of the magnetic field with respect to
the plane of the disk plays a role its affect should be visible in the
amplitude of the kHz QPOs in Cygnus X-2 when these QPOs are detected
during different overall intensity levels, thus at different phases of
the 78 day long-term X-ray cycle.

{\it Note added in manuscript} After submission of this paper we
discovered also two simultaneous kHz QPOs in the Z source GX 340$+$0
at 348$^{+20}_{-16}$ Hz and 722\pp13 Hz (peak separation of 374\pp24
Hz, when the source was on the horizontal branch (Jonker et
al. 1997). The FWHM and rms amplitude (in the energy range 5.0--60
keV) were 114$^{+64}_{-39}$ Hz and 2.5$^{+0.5}_{-0.4}$\% for the
lower-frequency QPO, and 177$^{+58}_{-45}$ Hz and
3.8$^{+0.5}_{-0.4}$\% for the higher-frequency QPO.

\acknowledgments

This work was supported in part by the Netherlands Foundation for
Research in Astronomy (ASTRON) grant 781-76-017 and by NSF grant AST
96-18524. B.V. (NAG 5-3340), F.K.L (NAG 5-2925), J.v.P. (NAG 5-3269,
NAG 5-3271) and W.H.G.L.  acknowledge support from the United States
National Aeronautics and Space Administration. We thank the RXTE
Science Operations Facility, and especially John Cannizzo, for rapidly
providing the real-time data used in this analysis.

\clearpage
\clearpage
\clearpage

\clearpage

\end{document}